\documentclass[12pt,preprint]{aastex}
\usepackage{graphicx}
\shorttitle{Coronal plasma density fluctuations}
\shortauthors{Telloni et al.}
\begin{document}
\title{Detection of plasma fluctuations in white-light images of the outer solar corona: investigation of the spatial and temporal evolution}
\author{D. Telloni\altaffilmark{1}, R. Ventura\altaffilmark{2}, P. Romano\altaffilmark{2}, D. Spadaro\altaffilmark{2} and E. Antonucci\altaffilmark{1}}
\altaffiltext{1}{National Institute for Astrophysics (INAF), Astrophysical Observatory of Torino, Via Osservatorio 20, 10025 Pino Torinese, Italy}
\altaffiltext{2}{National Institute for Astrophysics (INAF), Astrophysical Observatory of Catania, Via S. Sofia 78, 95123 Catania, Italy}
\begin{abstract}
This work focus on the first results on the identification and characterization of periodic plasma density fluctuations in the outer corona, observed in STEREO-A COR1 white-light image time series. A 2D reconstruction of the spatial distribution and temporal evolution of the coronal fluctuation power has been performed over the whole plane of the sky, from 1.4 to 4.0 $R_{\odot}$. The adopted diagnostic tool is based on wavelet transforms. This technique, with respect to the standard Fourier analysis, has the advantage of localizing non-persistent fluctuating features and exploring the variations of the relating wavelet power in both space and time. The map of the variance of the coronal brightness clearly outlines intermittent, spatially coherent fluctuating features, localized along, or adjacent to, the strongest magnetic field lines. In most cases they do not correspond to the coronal structures visible in the brightness maps. The results obtained provide a scenario in which the solar corona shows quasi-periodic, non-stationary density variations characterized by a wide range of temporal and spatial scales and strongly confined by the magnetic field topology. In addition, structures fluctuating with larger power are larger in size and evolve more slowly. The characteristic periodicities of the fluctuations are comparable to their lifetimes. This suggests either that plasma fluctuations lasting only one or two wave periods and initially characterized by a single dominant periodicity, rapidly decay into a turbulent mixed flow via nonlinear interactions with other plasma modes, or that they are damped by thermal conduction. The periodic non-stationary coronal fluctuations outlined by the closed field lines at low and mid latitudes might be associated with the existence of slow standing magneto-acoustic waves excited by the convective supergranular motion. The fluctuating ray-like structures observed along open field lines appear to be linked either to the intermittent nature of the processes underlying the generation of magnetic reconnection in the polar regions or to oscillatory transverse displacements of the coronal ray itself.
\end{abstract}
\keywords{methods: data analysis---magnetohydrodynamics (MHD)---Sun: corona---Sun: magnetic topology---Sun: oscillations---waves}
\section{Introduction}
MagnetoHydroDynamic (MHD) oscillations and waves in the coronal plasma are of great interest in solar physics due to the crucial role that these processes might play in coronal heating and solar wind acceleration \citep[e.g.][]{cranmer1999,walsh2003,ofman2004}. Coronal waves are also important as a probe for MHD coronal seismology \citep[e.g.][]{uchida1970,roberts1984,nakariakov2005}: the comparison of the observed properties of these phenomena, such as periods, wavelengths, amplitudes, spatial and temporal scales and time evolution, with theoretical models may lead to cast light on physical quantities of the solar corona, such as the magnetic field and the transport coefficients, not yet accurately known from the observational point of view.

Periodic and quasi-periodic phenomena in the solar corona were first observed with ground-based radio instrumentation in the 80s \citep{aschwanden1987}. The launch of SOHO \citep[SOlar and Heliospheric Observatory,][]{domingo1995}, in 1995, and TRACE \citep[Transition Region And Coronal Explorer,][]{handy1999}, in 1998, opened a new era in the investigation of coronal waves and transients. The highly improved spatial and temporal resolution of the new instrumentation carried by these spacecraft allowed the identification, almost ubiquitously in the corona, of a large variety of MHD waves: standing fast kink disturbances, fast sausage and slow longitudinal acoustic modes, propagating slow and fast waves, and so on. They permeate individual structures of the inner corona, such as loops, polar plumes and helmet streamers, with different spatial and temporal scales and periods ranging from a few seconds to several minutes. Longer period ($8-27\,h$) oscillations have been observed with EIT/SOHO \citep[Extreme ultraviolet Imaging Telescope,][]{delaboudiniere1995} in the inner corona in a large-scale EUV filament \citep{foullon2004}. UVCS/SOHO \citep[UltraViolet Coronagraph Spectrometer,][]{kohl1995} has allowed to discover density fluctuations in the outer corona, in the region where the solar wind is accelerated. Slow magneto-acoustic waves with periods of a few minutes ($6-10\,min$) have been detected in a coronal hole between $1.5\,R_{\odot}$ and $2.1\,R_{\odot}$ by \citet{ofman2000} and \citet{morgan2004}. Large-scale density fluctuations with longer periods (from a few hours to a few days) have been discovered in both the fast wind at high latitude at $2.1\,R_{\odot}$ \citep{bemporad2008,telloni2009b} and in the slow wind at mid and low latitudes at $1.7\,R_{\odot}$ \citep{telloni2009a}. The density fluctuations found in the slow wind can be interpreted as due to the existence of inhomogeneities locally generated by nonlinear interactions and carried out by the expanding coronal plasma. In alternative they can be also ascribed to outward propagating magneto-acoustic waves characterized by wavelengths compatible with the supergranular scale in the convection zone.

In spite of the wealth of the existing observations, the investigation on coronal plasma fluctuations is still fragmentary and affected by heavy limitations. Up to now, only the dominant temporal scales of periodic and quasi-periodic variability have been detected, and in general information on their temporal evolution is still lacking. Moreover, due to the limited instantaneous Field of View (FOV) of the spectrometers used in the previous studies, only a small fraction of the Plane Of the Sky (POS) has been investigated at a given time. This did not allow us to derive the spatial distribution of the plasma fluctuations.

Aim of this paper is to overcome these major limitations in the study of the periodic and quasi-periodic variability of the outer corona, by exploiting the improved observational capabilities of STEREO-A COR1 \citep{thompson2003,howard2008,kaiser2008}, launched in 2006, and by adopting a diagnostic technique apt to provide both spatial and temporal information on coronal plasma fluctuations. This is achieved by analyzing long-term time series of the total brightness derived from the high-cadence white-light images obtained with COR1 and by adopting the wavelet transforms as analysis technique, which allow us to separate spatial and temporal variations. The detailed study of the spatiotemporal evolution of the non-stationary plasma fluctuations, observed at different timescales, allows a 2D reconstruction of the wavelet power which maps the outer corona from 1.4 to $4.0\,R_{\odot}$ in a systematic manner and over a time interval of 9 days. This approach allows us to investigate the correlation between coronal periodic and quasi-periodic phenomena and magnetic structures and to attempt an identification of the origin, excitation and dissipation mechanisms of the detected variability.

The layout of the paper is as follows: description of the observations (\S{} 2), description of the analysis of the data and presentation of the results (\S{} 3), discussion (\S{} 4) and interpretation (\S{} 5) of the results, conclusions (\S{} 6). The appendix reports methodological details of the wavelet technique.
\section{Observations}
The present analysis is based on white-light images of the outer solar corona obtained during the minimum of solar activity in 2008 with the coronagraph COR1-A, which is part of the SECCHI (Sun-Earth Connection Coronal and Heliospheric Investigation) instrument suite aboard the 'ahead' spacecraft of the twin STEREO (Solar Terrestrial RElations Observatory) satellites. COR1-A observes the solar corona from 1.4 to $4.0\,R_{\odot}$, providing 1024$\times$1024 pixel images with a spatial resolution of $7.5\,arcsec$ ($\sim0.008\,R_{\odot}$) per pixel.

The entire set of observations consists of $N=2592$ exposures acquired with a cadence of $5\,min$, in the 9 days interval between April 14, 2008 at 00:05 UT and April 22, 2008 at 23:55 UT. During the selected observational run, the Sun was in an extremely quiet period with very few sunspots, small active regions and no evidence of eruptive phenomena.

\begin{figure*}
  \centering
  \includegraphics[width=\hsize]{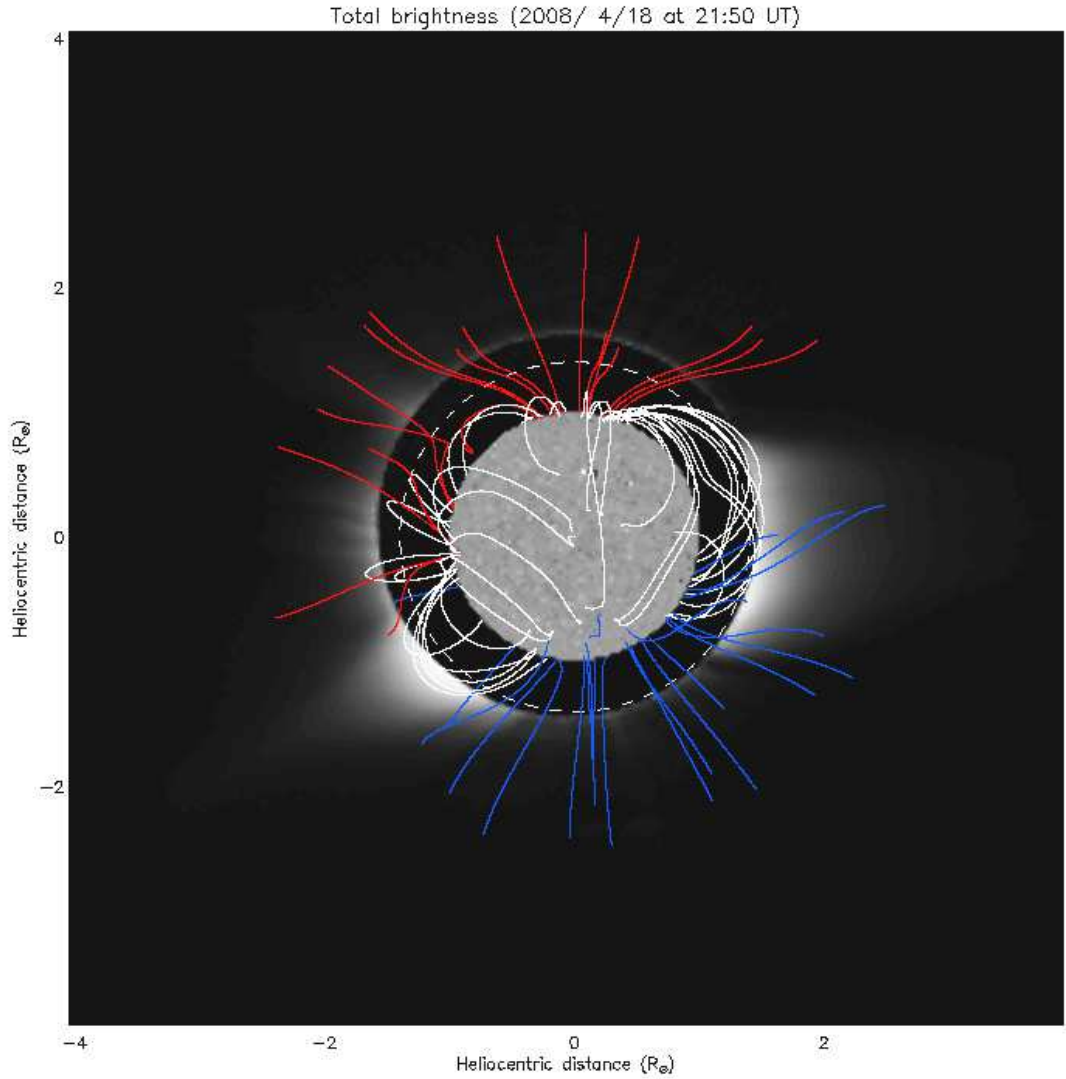}
  \caption{Total brightness of the outer solar corona, from 1.4 to $4.0\,R_{\odot}$, derived from the COR1-A observations obtained on April 18, 2008 at 21:50 UT; the extrapolated magnetic field lines and the SOHO/MDI disk magnetogram, obtained on April 19, 2008 at 00:04 UT as seen from STEREO-A are also shown; the white dashed line indicates the edge of the STEREO occulter.}
  \label{fig:total_brightness}
\end{figure*}

Continuous temporal series of white-light total brightness are derived for each pixel of the COR1-A images by using the IDL routine SECCHI\_PREP in the SolarSoft library. This routine allows also the calibration of the data and the subtraction of the background. The total brightness is computed in mean solar brightness $B_{\odot}$ units by combining the intensity radiation $I$ of a rapid polarization sequence triplet measured at polarizer positions 0, 120, 240 degrees: $B=\frac{2}{3}(I_{0}+I_{120}+I_{240})$ \citep{billings1966}. As an example, Fig. \ref{fig:total_brightness} shows the total brightness of the solar corona on April 18, 2008 at 21:50 UT; the SOHO/MDI \citep[Michelson Doppler Imager,][]{scherrer1995} magnetogram (as seen from the STEREO-A satellite on April 19, 2008 at 00:04 UT) is superimposed to the white-light image. In the same figure, the magnetic configuration of the corona is outlined by extrapolating the coronal magnetic field lines according to the model developed by \citet{schrijver2003} (employed in the SolarSoft package named PFSS); this model is based on the evolving full-Sun Carrington maps of the photospheric magnetic field measured by the SOHO/MDI instrument. The coronal magnetic field is extrapolated from the photospheric one via the Potential-Field Source-Surface (PFSS) approximation, in which the field is assumed potential in the coronal volume between the photosphere and a spherical source surface located at a height of $2.5\,R_{\odot}$. Since the coronal field models are provided at six-hour cadence by the online database of PFSS, Fig. \ref{fig:total_brightness} shows the magnetic configuration closest in time to the total brightness frame of April 18, 2008 at 21:50 UT. The extrapolations have been generated taking into account the instantaneous line of sight of STEREO-A by considering the Carrington latitude and longitude values at each time.

\section{Analysis and results}
In this paper we adopt the wavelet analysis with the aim to identify periodic and quasi-periodic density fluctuations over the whole outer solar corona and to explore their fine spatial structure and temporal evolution at different timescales. In recent years, this technique, applied to observations of the inner corona, led to the detection of periodic phenomena such as: i) outward propagating oscillations in an active region bright loop with periods of $3-7\,min$ \citep{demoortel2000}, ii) sporadic oscillations with periods of $7-8\,min$ in the quiet corona \citep{morgan2004} and iii) coherent fluctuations in a weak magnetic network region \citep{bloomfield2004}.

The wavelet analysis is in fact a powerful tool for analyzing non-stationary time series or time series with expected localized variations of power. By decomposing a time series into the time-frequency space, it is possible to determine both the dominant frequencies/periodicities of the variability and the related power modulation in time \citep{torrence1998}. The main benefit of the wavelet over the Fourier analysis resides in the capability of identifying the periodic components of non-stationary signals. By combining the results obtained at multiple locations, the spatial distribution and temporal variability of a given data set can be reconstructed. Details on the wavelet formalism and methodology, with particular attention to the application to the present analysis, are reported in the Appendix.

We applied the wavelet transform to the 9-days time series of the total brightness derived for each pixel of the COR1-A images. Trends affecting the temporal series in this interval have been previously removed; the effect of the solar rotation is filtered out by applying a Fourier highpass band filter with a cutoff frequency of $2.8\times10^{-5}\,Hz$, by taking into account the coronal rotation rate derived by \citet{giordano2008}. The long uninterrupted time series and the short temporal sampling, $5\,min$, allows the detection of periodicities ranging from $10\,min$ (the Nyquist-Shannon period) up to $10\,h$, the cutoff timescale of the highpass filter.

\begin{figure*}
  \centering
  \includegraphics[height=\hsize]{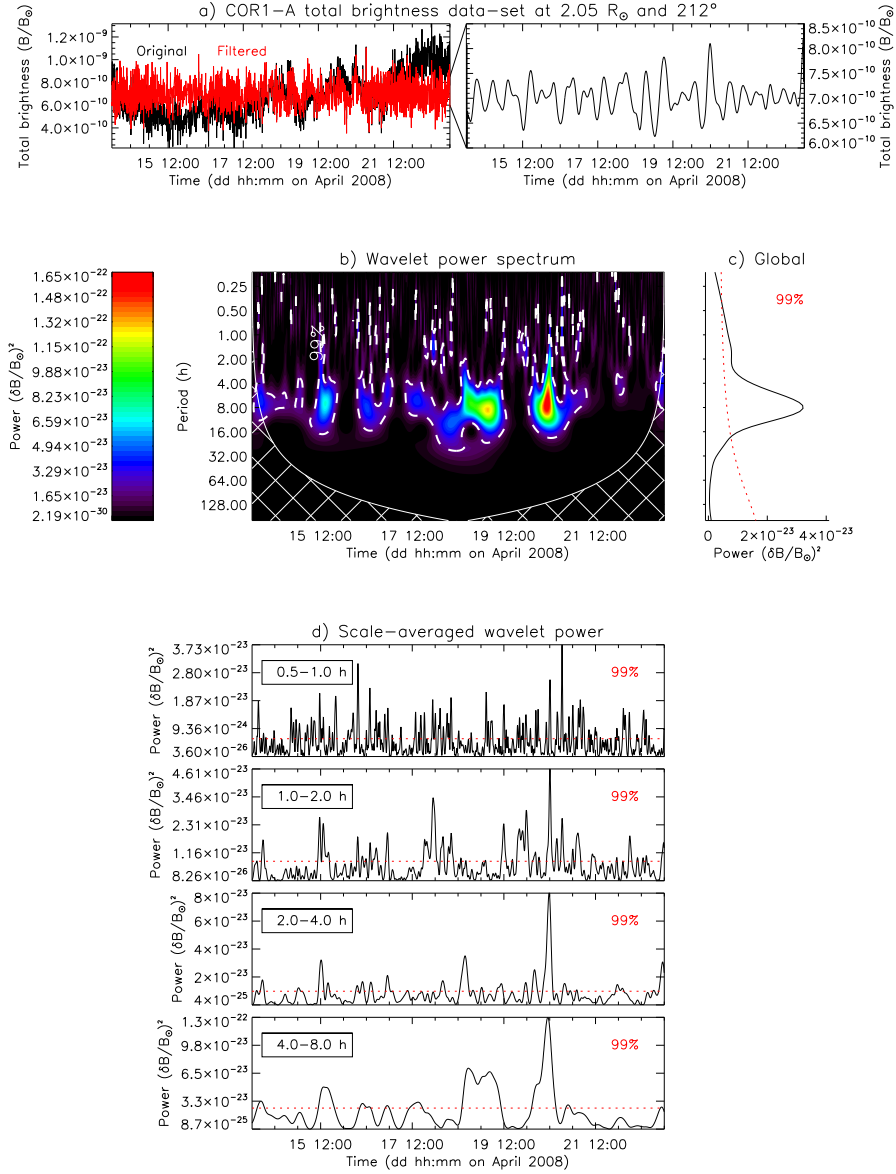}
  \caption{(a) \emph{left}: Coronal total brightness detected on April 18, 2008 at $2.05\,R_{\odot}$ and $212^{\circ}$ counterclockwise from the North pole (black curve); the red curve has been obtained by filtering out any trend, including that due to the solar rotation, from the original time series (black curve). (a) \emph{right}: Gaussian smoothing of the filtered time series (red curve on the left panel) which acts as lowpass filter for periods smaller than $2.5\,h$; note the finer scale resolution of the ordinate axis. (b) Wavelet power spectrum of the filtered time series, using the Paul wavelet function; the ordinate is the Fourier period $\tau$ in $h$; the abscissa is the time of observation; the continuous line is the Cone Of Influence (COI) and delimits a cross-hatched area where the power spectrum is not significant, whilst the thick dashed contours delimit areas where the wavelet power is significant with a 99\% statistical confidence, for a white-noise process. (c) Wavelet power spectrum averaged over the entire observational interval for the dataset reported in (a) (i.e. global wavelet spectrum); the red dashed line is the 99\% statistical confidence for a white-noise process. (d) Averages of the wavelet power over the $0.5-1\,h$, $1-2\,h$, $2-4\,h$ and $4-8\,h$ periodicity bands (from top to bottom, respectively); the red dashed lines are the corresponding 99\% confidence levels.}
  \label{fig:wavelet_power_spectrum_paul}
\end{figure*}

Since the Sun was very quiet during the observational interval, without any evidence of eruptive or transient phenomena, the data are smoothly varying without abrupt jumps or sharp peaks. As an example, the left panel of Fig. \ref{fig:wavelet_power_spectrum_paul}a shows the total brightness measured at $2.05\,R_{\odot}$ and $212^{\circ}$ counterclockwise from the North pole (which refers to a ray-like structure, see \S{} 3.1) before (black curve) and after (red curve) the filtering process. The right panel of Fig. \ref{fig:wavelet_power_spectrum_paul}a shows the result of a Gaussian smoothing of the red curve which acts as a lowpass filter for period smaller than $2.5\,h$ and allows us to cast a preliminary rough glance at the total brightness variability behavior. Highly unstable fluctuations whose amplitudes vary even significantly with time (strong impulses alternate with phases of damped amplitude) are present all along the observational run. This behavior is noticed in arch-shaped structures as well (see \S{} 3.1). Figure \ref{fig:wavelet_power_spectrum_paul}b shows the wavelet power spectrum derived from the filtered brightness time series reported as a red curve in the left panel of Fig. \ref{fig:wavelet_power_spectrum_paul}a, by using the Paul mother wavelet (see the appendix). The ordinate is the Fourier period $\tau$, corresponding to the scale $s$ of the wavelet, the abscissa the time of observation. The contour plot indicates the power of the total brightness fluctuations derived by means of the wavelet transforms, that is the variance of the total brightness time series detected by STEREO-A COR1. The cross-hatched area below the Cone Of Influence (COI) line (continuous line in Fig. \ref{fig:wavelet_power_spectrum_paul}b) is the region of the wavelet spectrum where edge effects, due to the finite-length of the time series, are significant. The shorter periodicities found above the COI line are due to the non-stationary periodic signals present in the data. The thick dashed contours delimit regions where the wavelet power is significant, that is they identify real periodic signals at a confidence level of 99\%, as determined via a 2 degree of freedom $\chi^{2}$ distribution \citep[see][and references therein as well as the appendix]{torrence1998}.

As shown in Fig. \ref{fig:wavelet_power_spectrum_paul}b, most of the power is concentrated in the $4-8\,h$ periodicity range. In this band its value exhibits significant changes, with a peak on April 20, 2008 at about 12:00 UT. As expected, negligible power is found at periods larger than ten hours, where frequency contributions have been filtered out. Hence, we are confident that the high-power features identified during the observational run are not to be ascribed, for instance, to the passage of coronal features across the instantaneous FOV during the solar rotation.

A vertical slice through the wavelet plot of Fig. \ref{fig:wavelet_power_spectrum_paul}b is a measure of the spectrum at a given instant. The average of the local wavelet power spectra over the observational interval gives the global wavelet spectrum (as defined in Eq. \ref{eq:global_wavelet_spectrum}) shown in Fig. \ref{fig:wavelet_power_spectrum_paul}c. It is worth noting that the global wavelet spectrum approaches the results that can be obtained with the Fourier spectrum of the time series \citep{percival1995}. The global spectrum is above the 99\% confidence level (red dotted line in Fig. \ref{fig:wavelet_power_spectrum_paul}c) in correspondence with the periodicities in the $4-8\,h$ band where the highest power is found.

In order to examine how the wavelet power change throughout the observational run, the power spectrum has been divided in four different periodicity bands ($0.5-1\,h$, $1-2\,h$, $2-4\,h$ and $4-8\,h$). For each band the average wavelet power, that is the average of the power over all the periodicities within a selected range $[s_{1}$ -- $s_{2}]$, has been computed (Eq. \ref{eq:scale_averaged_wavelet_power}). This quantity is a measure of the average variance of the total brightness as a function of time. The total brightness power exhibits significant modulations with timescales that are of the same order of the periodicities considered: the primary peaks in each band are one or two periods wide. Moreover, the longer the periods considered the slower the temporal evolution of the total brightness variance (Fig. \ref{fig:wavelet_power_spectrum_paul}d). We will discuss a possible physical interpretation of this temporally intermittent behavior in \S{} 4 and \S{} 5.

\begin{figure*}
  \centering
  \includegraphics[height=\hsize]{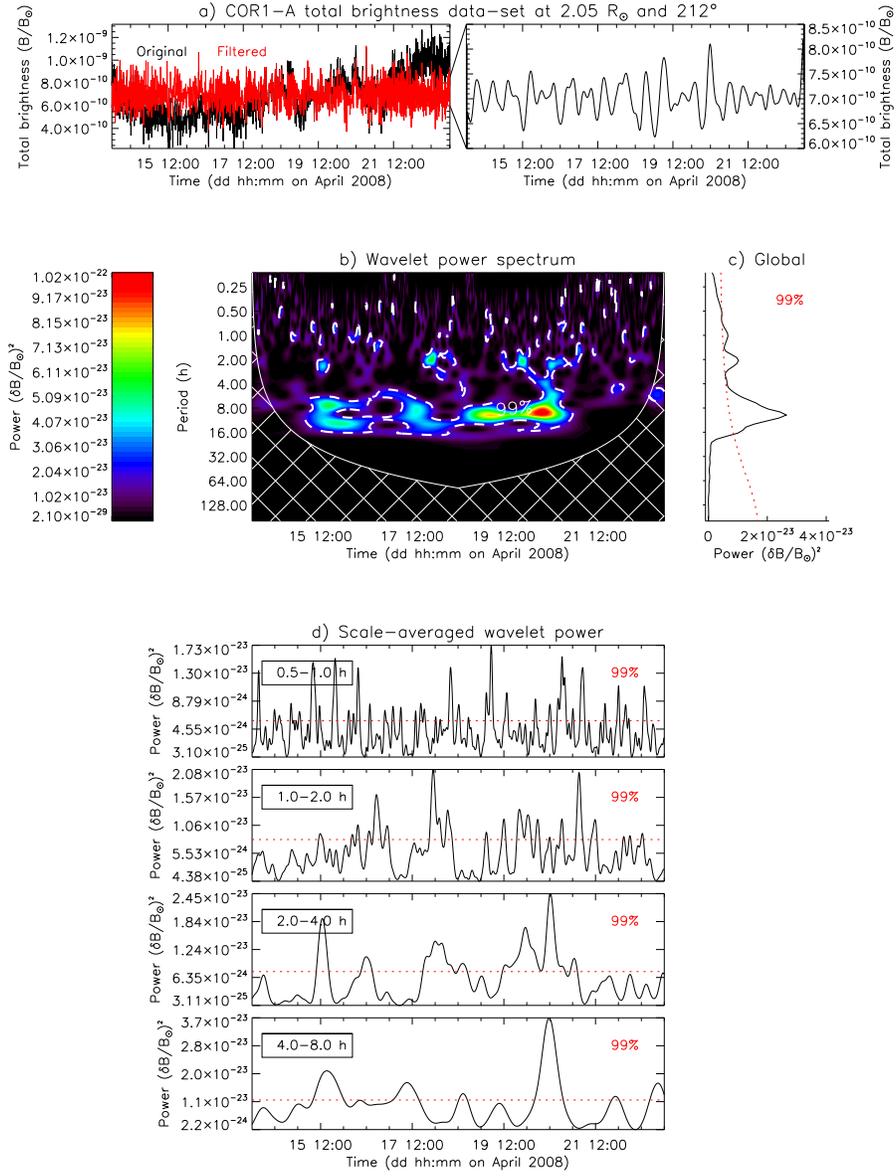}
  \caption{Same as Fig. \ref{fig:wavelet_power_spectrum_paul}, in the case of the analysis performed with the Morlet wavelet.}
  \label{fig:wavelet_power_spectrum_morlet}
\end{figure*}

Since the results of the wavelet analysis is strongly dependent on the choice of the wavelet function (see the appendix), the present results (Fig. \ref{fig:wavelet_power_spectrum_paul}b) are compared with those derived using the Morlet wavelet (Fig. \ref{fig:wavelet_power_spectrum_morlet}b).

The same features appear to be significant in both wavelet power spectra, approximately at the same time-periodicity locations and with comparable power. However, since the Paul mother wavelet is more sharply defined in time than the Morlet function (Fig. \ref{fig:paul_morlet_wavelet}), the peaks in the power spectrum shown in Fig. \ref{fig:wavelet_power_spectrum_paul}b are sharper in the time direction than those derived in Fig. \ref{fig:wavelet_power_spectrum_morlet}b, thus indicating a better time resolution in the case of the Paul wavelet function. On the other hand, since the Morlet mother wavelet is narrower in the periodicity space than the Paul function (Fig. \ref{fig:paul_morlet_wavelet}), the same peaks appear less elongated and better defined in the period direction, thus indicating that a better frequency resolution is obtained if the Morlet function is applied (Figs. \ref{fig:wavelet_power_spectrum_morlet}b and \ref{fig:wavelet_power_spectrum_paul}b).

The wavelet transforms based on the Paul mother wavelet (characterized by few zero crossings, Fig. \ref{fig:paul_morlet_wavelet}) can in principle identify many types of signatures in the power spectrum, such as sharp variations in the input data like simple rounded impulses, and not just waves. Nevertheless, by taking into account that the same significant high-power features are found in both the Paul and Morlet analyses and that the Morlet function is more suitable to detect signatures of periodic and quasi-periodic fluctuations (since it contains more "waves" than the Paul function, Fig. \ref{fig:paul_morlet_wavelet}), it is possible to infer that the signals detected by COR1-A might be not simply variable.

Thus, since the main aim of this study is to localize in time and in space any quasi-periodic feature associated with expected non-stationary processes, the Paul mother wavelet, which assures a better time localization than the Morlet function, has been used throughout the analysis reported hereinafter.

In summary, Figs. \ref{fig:wavelet_power_spectrum_paul}b,c,d show that significant non-stationary variability exists over a broad range of periodicities, extending from 0.5 up to $8\,h$; the strongest fluctuations are found in the $4-8\,h$ band. The observed variability persists throughout the observational interval. The temporal evolution of the wavelet power is strictly related to the periodicities involved: longer periodicities are characterized by a more slowly evolving wavelet power.
\subsection{2D reconstruction of the wavelet power}
\begin{figure*}
  \centering
  \includegraphics[width=\hsize]{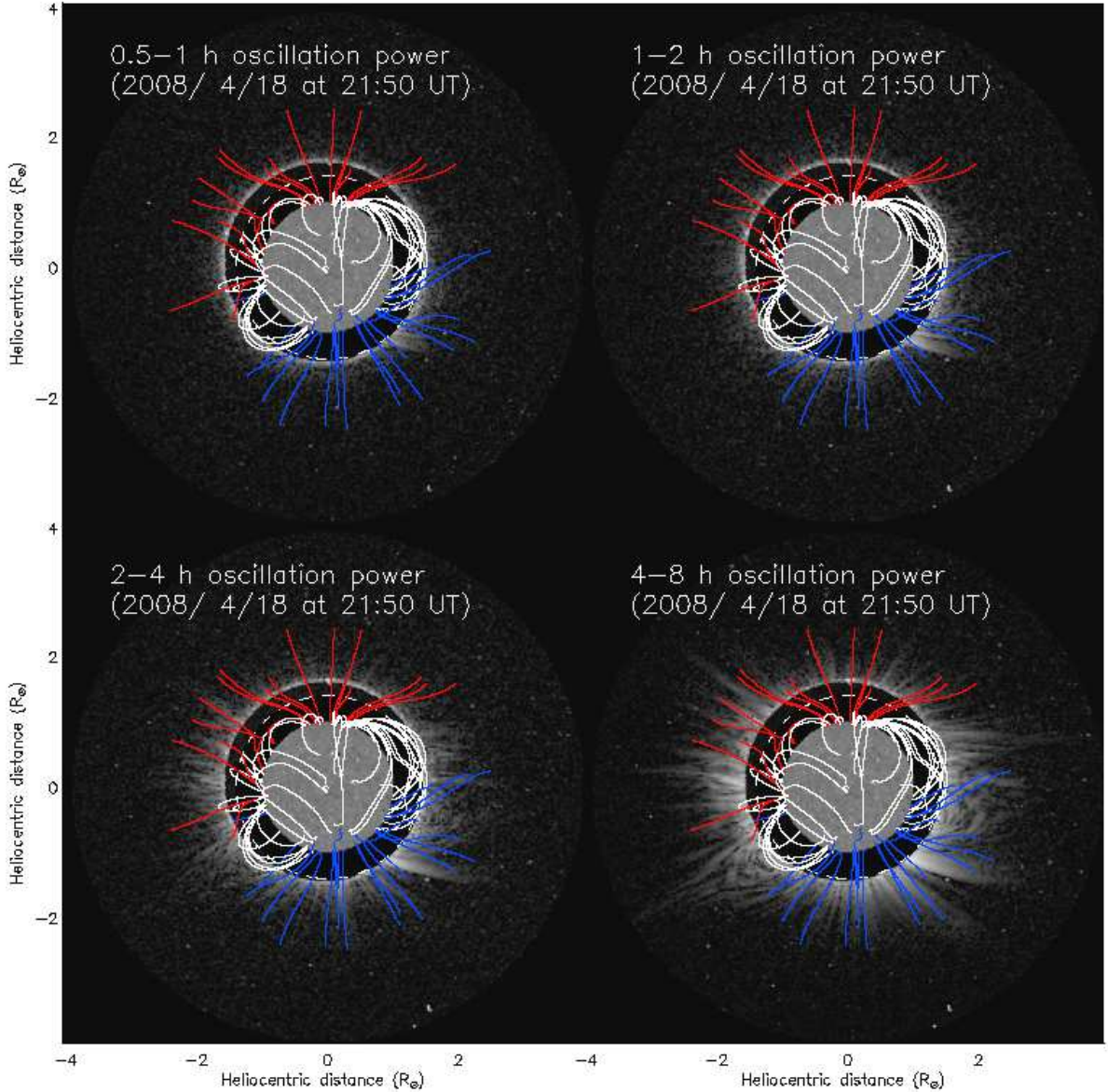}
  \caption{2D maps of the total brightness fluctuation power averaged over the $0.5-1\,h$, $1-2\,h$, $2-4\,h$ and $4-8\,h$ periodicity bands respectively, obtained in the FOV of COR1-A on April 18, 2008 at 21:50 UT; the extrapolated magnetic field lines and the SOHO/MDI disk magnetogram, as seen from STEREO-A on April 19, 2008 at 00:04 UT, are also shown; the plotted average variance of the total brightness is significantly above the 99\% confidence for the assumed white noise; the dashed white line indicates the edge of the STEREO occulter.}
  \label{fig:oscillation_power}
\end{figure*}

The spatial and temporal evolution of the total brightness fluctuation power is obtained by computing the averages (Eq. \ref{eq:scale_averaged_wavelet_power}) as functions of time, pixel by pixel, for the four selected periodicity bands. The results have been used to construct four different movies (available as online material in the electronic version of this paper), which show non-stationary, highly unstable fluctuating features in the entire FOV observed by STEREO-A COR1, appearing as arch-shaped (at low and mid latitudes) or ray-like (at higher latitudes) structures. The lifetimes of such features are comparable to the characteristic periodicities of each band. A single frame of each movie at a given instant is derived by combining into a two-dimensional contour plot the values of the average wavelet power. Figure \ref{fig:oscillation_power} shows the 2D maps of the wavelet power in the four different periodicity bands, at the same instant as in Fig. \ref{fig:total_brightness}; the extrapolated magnetic field lines (April 19, 2008 at 00:04 UT) are superimposed to the 2D maps. The figure also shows the SOHO/MDI disk magnetogram (as seen from the STEREO-A satellite) at that epoch. The power is plotted when exceeding the 99\% confidence level (derived as specified in the appendix). The features identifiable in the different bands are significant non-stationary fluctuating structures, with power well above the background level. The background is mainly due to random fluctuations of the total brightness of the quiet solar corona.

The features corresponding to power enhancements are often but not always localized along the magnetic field lines (Fig. \ref{fig:oscillation_power}). Moreover, they do not correspond to bright resolved structures with sharp edges in the total brightness images of the solar corona. The power of the bright features, shown in the Fig. \ref{fig:oscillation_power}, is about ten times larger than that of the dimmer regions (Fig. \ref{fig:wavelet_power_spectrum_paul}d). For instance, the power associated with the bright polar rays evident in the $4-8\,h$ band is about $2-4\times10^{-23}\,(\delta B/B_{\odot})^{2}$, whilst that of the dim regions between bright rays is $1-2\times10^{-24}\,(\delta B/B_{\odot})^{2}$. Similar ratios are found for the wavelet power in the bright equatorial arches compared to that derived for the dark cavities. The periods of the embedded fluctuations are related to the thickness ($d_{\perp}$) and the length ($L$) of the arch-shaped and ray-like features, determining the spatial scales of the fluctuating structures; that is, periodicities are longer for larger spatial scales. As an example, in the $4-8\,h$ band, the thickness of the fluctuating structures is $d_{\perp}\sim0.01\,R_{\odot}=7\,Mm$, while the extrapolated lengths (down to the solar disk) are $L\sim2.9\,R_{\odot}=2.03\times10^{3}\,Mm$ and $L\sim1.4\,R_{\odot}=980\,Mm$ for the arches and rays, respectively.
\subsection{Spatial and temporal profiles of the fluctuation power}
In order to investigate in detail how the power varies and evolves along the bright non-stationary fluctuating features, spatial and temporal profiles of the power of the total brightness are derived as a function of a linear coordinate along arches and rays (Fig. \ref{fig:oscillation_power}). Since these structures are more evident for periods of $4-8\,h$, hereafter we focus our attention on this periodicity band.

\begin{figure*}
  \centering
  \includegraphics[height=\hsize,angle=90]{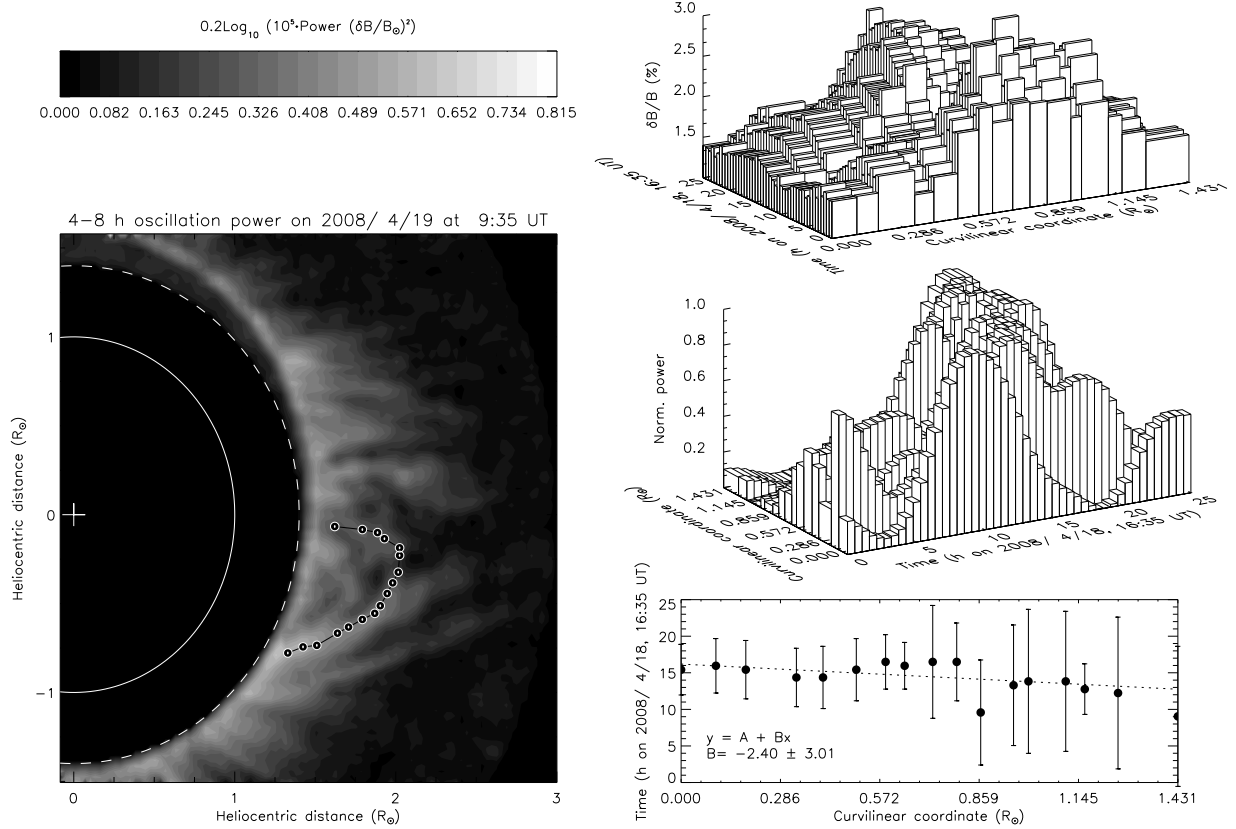}
  \caption{\emph{Left}: Wavelet power averaged over the $4-8\,h$ periodicity band, inferred in the western hemisphere of the COR1-A FOV on April 19, 2008 at 09:35 UT; the dots outline an arch-shaped fluctuating structure; the continuous white line indicates the solar disk limb, whilst the dashed one the edge of the STEREO occulter. \emph{Top right}: 3D histogram showing the spatial profiles of the variability amplitude relative to the instantaneous total brightness referring to different points along the arch, at different times (one every about two hours), from April 18, 2008 at 16:35 UT to April 19, 2008 at 17:35 UT. \emph{Middle right}: 3D histogram showing the temporal profiles of the power of the fluctuations inferred along the arch, in the same time interval as in the top right panel. \emph{Bottom right}: Time instants when the wavelet power of the different temporal profiles reaches its maximum as a function of the curvilinear coordinate along the arch-shaped structure; the linear fit is shown as a dotted line.}
  \label{fig:spatial_temporal_profiles_loop}
\end{figure*}

The left panel of Fig. \ref{fig:spatial_temporal_profiles_loop} shows the power averaged over the $4-8\,h$ band, in the western hemisphere of the COR1-A FOV on April 19, 2008 at 09:35 UT. There is clear evidence for a system of two encapsulated closed arches, protruding from the occulted coronal region, i.e. from $1.4\,R_{\odot}$, up to about $1.7\,R_{\odot}$ and $2.0\,R_{\odot}$, respectively. The arches are separated by a dark cavity. The dots outline the outer arch.

The spatial profiles of the fluctuation amplitude along the outer arch are plotted as a 3D histogram in the top right panel of Fig. \ref{fig:spatial_temporal_profiles_loop}. The x-axis represents the curvilinear coordinate defined as the incremental distance along the path from the southernmost visible part of the arch. The y-axis represents the time (in hours) starting on April 18, 2008 at 16:35 UT. The spatial behavior of the total brightness fluctuations along the arch is measured at each point by the percentage variation of their amplitude relative to the instantaneous total brightness (z-axis). Spatial profiles (along the x-axis) are obtained at a cadence of two hours in the period from April 18, 2008 at 16:35 UT to April 19, 2008 at 17:35 UT. The relative amplitude of the total brightness fluctuations increases along the arch, reaching its maximum of about 3\% around the cusp of the closed structure, a factor $\sim2$ higher than at the base. The trend of the spatial profiles is quasi-independent of the time.

The temporal profiles of the fluctuation power inferred at each point along the arch are plotted as a 3D histogram in the middle right panel of the same figure. The time interval over which the temporal profiles have been derived is much longer (more than $24\,h$) than the lifetime of the fluctuating feature, thus assuring that its entire temporal evolution is indeed monitored. The x-axis represents the time (in hours) starting on April 18, 2008 at 16:35 UT. The y-axis represents the curvilinear coordinate along the arch. The z-axis is the normalized wavelet power for each temporal profile. The total brightness fluctuations along the arch are typically characterized by a monotonic damping. This suggest that the fluctuating structures found in the corona are non-stationary in nature. The $1/e$-width of the main fluctuation at each point of the arch provides information on the lifetime of the fluctuating structure, which is slightly less than ten hours, that is, of the same order of the temporal scales examined in the band of $4-8\,h$. The power reaches its maximum on April 19, 2008 at about 07:35 UT almost simultaneously along the arch, except at the northernmost base where the maximum is reached slightly earlier. This fact could be due to a tilt of the fluctuating arch relative to the POS, which would produce a variable column depth along the LOS, thus modulating the temporal profiles of the wavelet power along the structure.

The bottom panel of Fig. \ref{fig:spatial_temporal_profiles_loop} shows the temporal and spatial coordinates of the power peak in the oscillating arch. The error bars are derived from the $1/e$-width of the corresponding temporal profiles. The linear coefficient of the fit shown as a dotted line, $B=2.40\pm3.01\,h\,R_{\odot}^{-1}$, is consistent, within the uncertainties, with a flat curve. This implies that the total brightness fluctuation detected reaches its maximum power fairly simultaneously along the arch, hence the fluctuations are coherent in phase.

\begin{figure*}
  \centering
  \includegraphics[height=\hsize,angle=90]{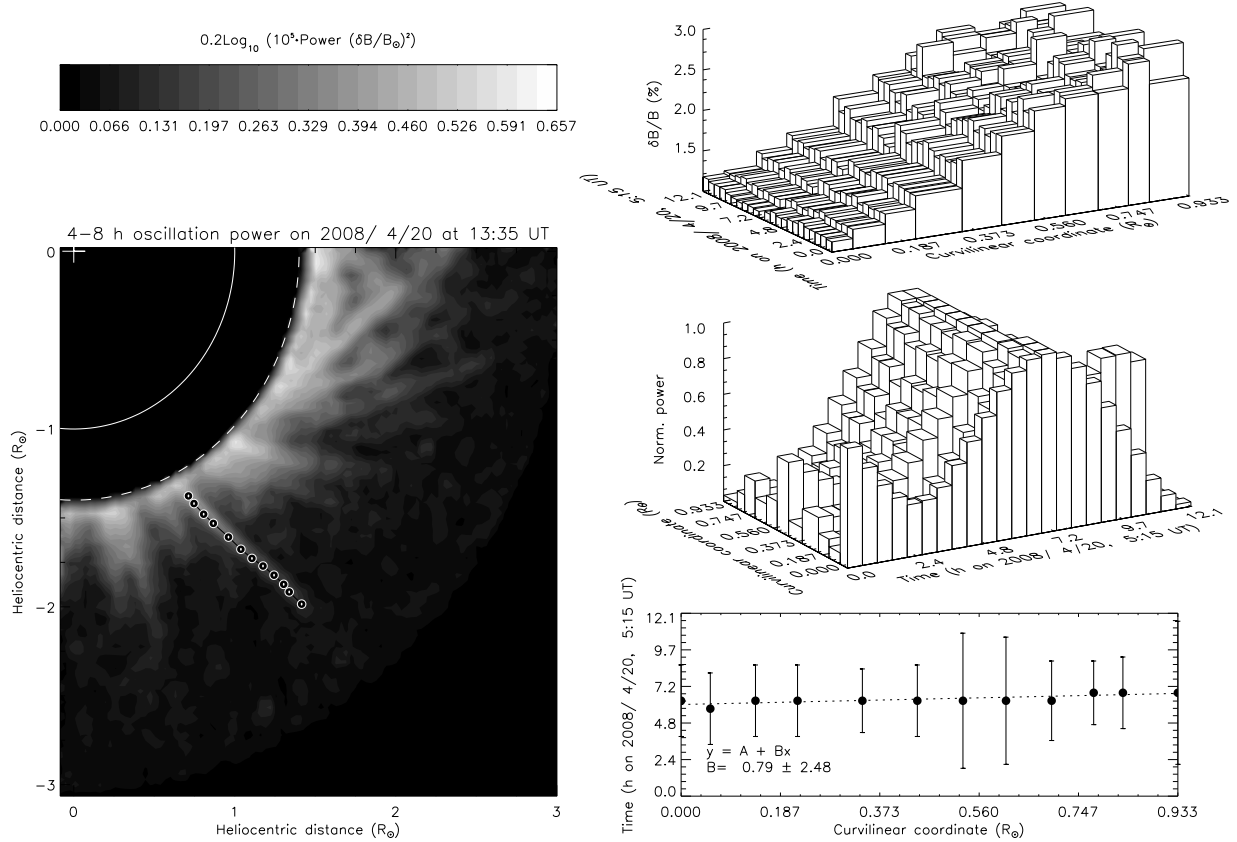}
  \caption{As in Fig. \ref{fig:spatial_temporal_profiles_loop} for a ray-like fluctuating structure (outlined by dots) in the south-western quadrant, on April 20, 2008 at 13:35 UT. The spatial and temporal profiles refer to the time interval from 05:15 to 17:20 UT on April 20, 2008, sampled every about one hour.}
  \label{fig:spatial_temporal_profiles_plume}
\end{figure*}

The same analysis has been also performed on a ray-like structure in the south-western hemisphere, shown in the left panel of Fig. \ref{fig:spatial_temporal_profiles_plume}, which shows the wavelet power averaged over the $4-8\,h$ band, on April 20, 2008 at 13:35 UT.

The spatial and temporal profiles of the wavelet power of the total brightness along the ray are shown as 3D histograms on the top and middle right panels of the figure. They are very similar to those obtained for the arch-like structure and can be summarized as follows: i) the fluctuations strengthen with distance along the ray: within $1\,R_{\odot}$ their amplitude, relative to the total brightness, increases by a factor of 2 up to 3\%; further out the structure is no longer observed; ii) the fluctuations are non-stationary and damped with a lifetime of about ten hours; the lifetime of the structure is of the same order of the periodicities of the embedded fluctuations, that is slightly less than ten hours; iii) the fluctuations along the ray-like structure are coherent in phase.
\section{Discussion}
Strong non-stationary fluctuating structures have been detected in the solar corona, as closed arches in the equatorial regions and ray-like features in the polar holes. They are often found in correspondence with the strongest lines of the coronal magnetic field as shown in Fig.\ref{fig:oscillation_power}. The observed variability can be ascribed to variations in the electron density of the emitting coronal plasma, since the white-light emission in corona is due to the Thomson scattering of the solar disk photons by the free electrons. The periodicities observed in the fluctuating corona cover a wide range of scales, from 0.5 to $8\,h$; the strongest signals are found in the $4-8\,h$ band, where the fluctuating structures have a characteristic transverse and longitudinal spatial scale of $\sim0.01\,R_{\odot}=7\,Mm$ and $\sim1.4-2.9\,R_{\odot}=980-2030\,Mm$, for the ray and arch features, respectively. These results confirm the findings of the first studies on density fluctuations in the outer solar corona by \citet{bemporad2008} and \citet{telloni2009a,telloni2009b}. In these papers, quasi-periodic oscillations with periodicities larger than a few hours were found in the Fourier power spectra; however, the analyses were much more limited in space and time, being performed with the data obtained with a spectrometer.

Notwithstanding we have ascertained that the total brightness variability is non-stationary in nature, it reveals a long-term persistence throughout the observational time interval of 9 days. This fact suggests that a non-transient, persistent mechanism of quasi-periodic nature, which not necessarily acts at coronal level, continuously drives density fluctuations in the solar corona. The fluctuations continuously originate and successively decay in a time that is of the same order of their characteristic temporal scales; this causes a modulation of the variance of the total brightness as seen in Figs. \ref{fig:wavelet_power_spectrum_paul}a and \ref{fig:wavelet_power_spectrum_paul}d and in the maps of the corona of Fig. \ref{fig:oscillation_power}. Such a behavior is clearly evident in the movies of the spatiotemporal evolution of the wavelet power (electronic version of this paper). It can be suggested that the density fluctuations, which are initially distinct and characterized by a single dominant periodicity, either rapidly decay into a turbulent mixed flow via nonlinear interactions with other plasma modes, or are damped by thermal conduction.

Furthermore, the results obtained in this study point to a tight correlation of the temporal evolution of the power of the plasma variability detected in this work and the spatial and temporal scales of the fluctuating coronal features. Longer periodicities are indeed typical of fluctuating structures of larger size and evolving more slowly. This reflects an overall characteristics of the Sun: larger-scale solar structures have characteristic timescales longer than those corresponding to smaller-scale structures. For instance, the strong correlation existing between the size and the time duration of the polar plumes \citep{llebaria2002} is well known. This behavior is as well evident in the movies (electronic version of this paper) that show the spatial and temporal evolution of the magnetic features, both arches and rays, in the corona from 1.4 to $4.0\,R_{\odot}$.

Finally, the analysis of the temporal evolution of the wavelet power along the non-stationary arch-shaped and ray-like structures (\S{} 3.2) reveals that the plasma fluctuations are highly coherent, since each point along the coronal magnetic structure fluctuates with the same phase. The phase coherence of the plasma fluctuations might indicate that they are likely to be generated simultaneously by the same driver mechanism.
\subsection{Assessment of solar rotation effects}
Even if the frequency contribution due to the solar rotation has been filtered out, in order to assure that the same volume of the corona is analyzed (as described in \S{} 3), it cannot be a priori excluded that, in observations performed at the limb, tangential discontinuities separating adjacent radial structures with different electron density (for instance, magnetic flux tubes co-rotating with the Sun), could mimic wave phenomena. During limb observations of the quiet corona, the maximum density along the line of sight is at the plane of the sky, due to its proximity to the disk. Thus a density variation caused by quasi-regularly spaced tangential discontinuities, due to the rotation of the Sun, induces a quasi-periodic signal in the coronal emission detectable at best when crossing the plane of the sky. We have tested this occurrence by evaluating the temporal scale of rotating long-lived structures, $\tau_{r}=d_{\perp}\frac{T_{r}}{(2\pi r)}$, where $d_{\perp}$ is the geometric thickness of the oscillating arch, perpendicular to the magnetic field lines, $T_{r}\sim27.5\,d$ is the rotation period observed in the outer corona \citep{giordano2008} and $r\sim1.5\,R_{\odot}$ is the average heliocentric distance of the fluctuating coronal arch. By assuming $d_{\perp}\sim0.01\,R_{\odot}$, that is the measured value in the $4-8\,h$ periodicity band, we obtain $\tau_{r}\lesssim1\,h$, which is significantly smaller than the periodicities characterizing the fluctuating arch. We have also addressed this issue by assessing the effect of the rotation on the temporal evolution of a coronal magnetic loop, formed by numerous spatially unresolved filamentary sub-structures \citep{klimchuk2009}. The loop fine structure has been modeled by considering a curved cylinder of longitudinal and transverse dimensions of $2.9\,R_{\odot}$ and $0.01\,R_{\odot}$, respectively (dimensions consistent with the length, $L$, and the thickness, $d_{\perp}$, of the observed arch-shaped structure), consisting of many sub-elements characterized by different densities. On the basis of this model we compute the coronal electron density as a function of time at the plane of the sky in a time interval of about one day. In the hypothesis that the randomly distributed sub-structures produce a correlated signal, the Fourier analysis of the synthetic coronal emission shows that the dominant mode of variability is consistent with the crossing timescale of the rotating arcade-system, i.e. $\sim1\,h$, whilst significantly less power is found at longer periods.

The effect of the solar rotation should be negligible in the case of polar structures which develop at high latitudes \citep{wilhelm2011}. For instance, the shape of the plumes on spatial scales of $\approx30\,arcsec$ is nearly invariant for hours to days \citep{waldmeier1955}. \citet{llebaria2002,llebaria2011} describe polar plumes as enduring structures showing brightness variations which might be ascribed to the solar rotation at timescales longer than about $10\,h$. The intrinsic variations of the plume brightness on scales of the order of $6\,h$, due probably to the intermittent nature of the generating process, are superimposed on this long-term variability. Therefore, in the $4-8\,h$ band, where, in a plume-like feature, we find significant variability, the effect of solar rotation should be negligible.

Before concluding that the arch-shaped power enhancements can be interpreted as due to intrinsic coronal density fluctuations or even more MHD waves, we have also to rule out another effect that could in principle mimic power in the wavelet maps, i.e. any possible spatial motion of an arch-shaped structure with clear edges. Quasi-periodic spatial displacements of coronal loops, detected by TRACE in the inner corona, have been reported in many papers \citep[e.g.][]{aschwanden1999}. They are commonly triggered by nearby flare events and are typically kink instabilities characterized by timescales of the order of a few minutes. Note, in this respect, that no flares or eruptive phenomena were detected during the examined observing run and, moreover, that the fluctuations found in this study are characterized by much longer timescales. Finally, power due to non-intrinsic variability such as that generated by a simple pulse of material injected in the arch with a duration appropriate to trigger the timescales of the detected fluctuations can be also ruled out on the basis of the high coherence of the signal found all over the characteristic fluctuation lifetimes.

The analysis shows that the non-stationary density fluctuations outlining closed and open coronal structures, such as arch-systems and ray-like features, can be ascribed to intrinsic brightness variations with possible MHD wave character. The compressive components of MHD waves can indeed perturb the ambient solar corona, modulate periodically or quasi-periodically the electron density and, in the end, generate plasma fluctuations, which can be observed by a white-light imaging coronagraph such as STEREO-A COR1.
\section{Interpretation}
\subsection{Arch-shaped structure fluctuations}
Theoretically various MHD waves are predicted in coronal loops. The dispersion relation for magneto-acoustic waves \citep[e.g.][]{priest1987} is satisfied by two kinds of waves: a slow-mode branch, with acoustic phase speeds, and a fast-mode branch, with Alfv$\acute{\textrm{e}}$n speeds. The class of the fast magneto-acoustic waves includes a symmetric sausage mode and an asymmetric kink mode. In the sausage mode, opposite sides of an arcade oscillate in anti-phase and the central axis remains undisturbed, while in the kink mode, the opposite sides and the central axis oscillate in phase. However, both modes drive plasma fluctuations characterized by the same phase along the arcade, as those observed in the coronal arch and detected in the present analysis (Fig. \ref{fig:spatial_temporal_profiles_loop}). Hence, these MHD modes, as well as slow-mode waves, can be considered physical candidates for the interpretation of the observed variability.
\subsubsection{Fast-mode waves}
Kink modes produce transverse amplitude oscillations. Oscillatory transverse displacements of coronal loops have been observed in the inner corona in EUV wavelengths with TRACE \citep[e.g.][]{aschwanden1999,aschwanden2002,nakariakov1999,schrijver1999,schrijver2002}. In the hypothesis that a standing kink-mode is responsible for the observed electron density fluctuations, the characteristic period $P_{kink}$ of such a mode at the fundamental harmonic can be derived, according to \citet{aschwanden2004}, as $P_{kink}\approx\frac{2L}{c_{\kappa}}\lesssim\frac{2L}{v_{A}}$, where $c_{\kappa}$ and $v_{A}$ are the kink and the Alfv$\acute{\textrm{e}}$n speed, respectively. In low-$\beta$ plasma conditions, where the thermal pressure is much smaller than the magnetic pressure, $c_{\kappa}\gtrsim v_{A}$, with $v_{A}\approx1000\,km\,s^{-1}$, for typical electron densities and magnetic fields in the coronal region of interest. For the observed arch with a measured length $L=2.03\times10^{3}\,Mm$, the characteristic period is $P_{kink}\approx1\,h$.

Hence, if the fluctuations detected in the coronal arch were due to standing kink waves, we would expect kink-mode periods, and in turn electron density fluctuations, of less than one hour. It turns out that, according to the analysis, the observed variability cannot be ascribed to a standing wave in the kink mode with fixed nodes at the footpoints.

Sausage MHD modes can be excluded as well because they are expected to be completely suppressed at small wavenumbers. In a coronal arch with length and cross section of the order of the observed ones ($L\simeq2.03\times10^{3}\,Mm$ and $d_{\perp}=0.01\,R_{\odot}$,) the dimensionless wavenumber is of the order of 0.005, much smaller than the cutoff wavenumber for typical coronal conditions, which falls into the range 0.8 -- 2.4 \citep{aschwanden2004}. As a consequence, the sausage type waves cannot explain the density fluctuations observed in the arch-shaped structure.
\subsubsection{Slow-mode waves}
Slow-mode oscillations in coronal loops have been imaged for the first time in the inner corona with the SUMER \citep[Solar Ultraviolet Measurement of Emitted Radiation,][]{wilhelm1995} spectrometer on SOHO, by measuring the Doppler shift and intensity variations of EUV lines \citep[e.g.][]{wang2002,wang2003a,wang2003b}. Slow-mode acoustic waves are essentially compressional, longitudinal sound waves. In coronal conditions, for typical temperatures of $T\sim1.5\,MK$, the sound speed $c_{s}\approx180\,km\,s^{-1}$ is generally much smaller than the Alfv$\acute{\textrm{e}}$n speed ($v_{A}\approx1000\,km\,s^{-1}$), that is $c_{s}/v_{A}\approx0.2$. Thus, the phase speed of slow magneto-acoustic waves in coronal loops, namely the tube speed $c_{T}$,

\begin{equation}
  c_{T}=\frac{c_{s}v_{A}}{\sqrt{c_{s}^{2}+v_{A}^{2}}}\approx c_{s}\qquad\textrm{for}\,c_{s}\ll v_{A}
  \label{eq:tube_speed}
\end{equation}

is close to the sound speed $c_{s}$ \citep[e.g.][]{priest1987}.

The period of a MHD slow-mode standing wave, having the two endpoints of a magnetic flux tube as fixed nodes, is the sound travel time of the fundamental harmonic, back and forth in the coronal loop, $P_{slow}=\frac{2L}{c_{T}}\approx\frac{2L}{c_{s}}$. Hence, the period of a slow magneto-acoustic standing wave, $P_{slow}\approx6\,h$, supported by the observed fluctuating arch structure falls in the $4-8\,h$ band where most of the power of the detected fluctuations is observed; thus, the fluctuation of the magnetic arch extending in the outer corona out to approximately $2\,R_{\odot}$ is consistent with a slow magneto-acoustic standing wave. Therefore, the MHD slow-mode standing waves are ideal candidates for the interpretation of the plasma fluctuations embedded in the coronal arch detected in the equatorial region of the quiet corona.

As far as the origin of the slow-mode wave is concerned, one possibility is that the variability is triggered by a pressure disturbance localized near one of the footpoints, which propagates as a slow-mode magneto-acoustic wave in longitudinal direction along the arch and is reflected at the opposite footpoint, thus generating a standing wave (a compression back and forth along the magnetic flux tube). A stochastic excitation by convective supergranular motions generates slow magneto-acoustic waves with a period which is consistent with those observed in the coronal fluctuating arch. The lifetimes of supergranular cells $\tau_{s}$ are ranging from 6 to $20\,h$ \citep{aigrain2004}; this interval is partially overlapping with the $4-8\,h$ band, where closed magnetic arches clearly exhibit the largest wavelet power. A further support to this interpretation is given by the value estimated for the wavelengths of slow-mode standing waves triggered by supergranulation, $\lambda_{s}=c_{T}\tau_{s}\approx c_{s}\tau_{s}\approx5.6\,R_{\odot}=3.89\times10^{3}\,Mm$. This wavelength is approximately twice the loop length $L$ of the fluctuating coronal arch, hence, it is consistent with a standing wave with fixed nodes at the footpoints of the magnetic flux tube. The present discussion suggests that supergranular convective motions very likely play a crucial role in the excitation of the slow MHD standing waves trapped in the large closed fluctuating magnetic structures extending in the outer corona. The length of the arch studied in this analysis is compatible with a spatial scale that can trap the standing wave, thus acting as a filter of standing waves with the appropriate wavenumber.

The suggestion that the driver of the standing waves along coronal arches can be identified with the supergranular convective motions is consistent with the results by \citet{telloni2009a}, where the characteristics of the coronal density fluctuations observed at $1.7\,R_{\odot}$ in low-latitude regions, such as the temporal and spatial scales, the degree of persistence and phase coherence, were all indicating a likely correlation between the variability of the solar corona and the photospheric dynamics induced by supergranular convection. However, \citet{telloni2009a} interpret the correlation between the dynamics of supergranulation and the coronal variability by assuming that structures of the supergranulation at the lower levels of the solar atmosphere is maintained in the outer corona; thus, coronal structures, crossing the POS during rotation, produce a periodic signal, whose characteristics depend on the temporal and spatial scales of the supergranulation. In this study, the link to convective motions is different, since, according to the interpretation of the present results, the stochastic motion of the supergranular cells acts as a driver for the excitation of slow magneto-acoustic waves which remain trapped in the arch structure as standing waves when their wavelengths are compatible with the loop length.
\subsubsection{Damping of MHD modes}
The MHD slow-mode standing waves, inferred from the density fluctuations of the arch observed in the quiet corona, show a strong damping effect, which has a decay time of the order of the fluctuation period, as shown in the middle right panel of Fig. \ref{fig:spatial_temporal_profiles_loop}. This rapid damping can be ascribed to processes such as thermal conduction, radiative losses and phase mixing. Thermal conduction and radiative losses are the major non-adiabatic cooling effects, which can contribute to the damping of the wave motion. For typical coronal electron densities ($n_{e}\sim1.2\times10^{7}\,cm^{-3}$) and temperatures ($T=1.5\,MK$), at the average heliocentric distance ($r\sim1.5\,R_{\odot}$) of the fluctuating arch of length $L\simeq2.03\times10^{3}\,Mm$, the cooling time due to thermal conduction from the corona to the chromosphere is of the order of $\tau_{cond}\approx5-6\,h$ \citep{priest1987}. Therefore, thermal conduction can be considered an important factor in damping standing slow magneto-acoustic waves trapped in the closed fluctuating coronal arch of Fig. \ref{fig:spatial_temporal_profiles_loop}, since the conductive cooling time is of the same order of the wave damping time, $4-8\,h$. The interpretation of thermal conduction as the main physical mechanism responsible for damping standing slow magneto-acoustic waves in magnetic loops is also supported by the MHD simulations performed by \citet{ofman2002} in order to interpret SUMER observations. Whilst, since the radiative cooling time, defined by the ratio of the thermal energy and the radiative cooling rate, is generally lower than one hour \citep{priest1987}, radiative losses are negligible in damping the MHD slow-modes, which have larger decay times.

Another mechanism that could substantially damp MHD slow-mode waves is phase mixing. Magneto-acoustic waves in a homogeneous medium, in density and temperature, propagate undamped and undisturbed because they are in perfect resonance with each other. However, coronal loops have large density and temperature gradients, especially near the footpoints where the standing waves are reflected. For instance, for MHD slow-mode waves, a density decrease along the coronal loop would increase the sound speed and decrease the wave periods along the magnetic flux tubes. Therefore, standing slow magneto-acoustic waves in a coronal loop might suffer intense phase mixing, during which modes of same frequency travel with different speed and then have different wavenumbers. As they propagate they become quickly mutually out of phase, leading to a destructive interference, and thus to wave dissipation \citep[e.g.][]{hood1997}. In other words, the slow MHD modes, which are initially distinct and of a single dominant periodicity, might rapidly decay into a turbulent mixed flow via nonlinear interactions with other reflected modes.

In conclusion both thermal conduction and phase mixing are viable mechanisms to explain the damping of the observed standing waves which manifest themselves in closed magnetic structures.
\subsection{Ray-like structure fluctuations}
In analogy to the proposed interpretation of the electron density fluctuations in coronal arches, we have evaluated the possibility that outward propagating slow MHD waves might be responsible as well for the non-stationary plasma fluctuations detected in a polar region along the magnetic plume-like feature shown in the left panel of Fig. \ref{fig:spatial_temporal_profiles_plume}, which might have an open magnetic configuration.

The density modulation due to propagating MHD waves would result in time-shifts of the temporal profiles of the wavelet power inferred along the polar ray. This is because nodes propagate with the wave. Hence, the density fluctuations are expected to be out of phase at different heliocentric altitudes along the open structure which might act as a channel for outward propagating waves. However, different part of a ray-like structure are observed to fluctuate in phase, reaching their maximum power quite simultaneously (middle and bottom right panels of Fig. \ref{fig:spatial_temporal_profiles_plume}). No evidence for magneto-acoustic wave propagation in the polar ray-like feature has been be found by studying the temporal cross-correlations (method described by \citet{duvall1993}) of the total brightness fluctuations observed at different locations along the structure. This clearly indicates that the plasma density fluctuations detected along the ray of Fig. \ref{fig:spatial_temporal_profiles_plume} have to be ascribed either to quasi-periodic motion forth and back across the POS of the entire structure or to intrinsic brightness variation of the polar ray itself, likely due to the intermittent nature of the process underlying its generation. As a matter of fact, although the magnetic configuration of solar polar regions is relative stable, the mechanism driving the formation of the actual ray structure might operate only intermittently \citep[e.g.][]{wang1998}.

Based on the characteristic timescales of the fluctuations detected in the present analysis, the main mechanism responsible for the temporal variations observed in the brightness of the polar features might be caused by magnetic reconnection which should be acting with comparable timescales. It has been suggested that magnetic reconnection of continuously emerging flux, guided by supergranular convection, could affect significantly the lifecycle of polar plumes, driving multiple hourly occurrences of the plume brightening \citep[e.g.][]{wang1998,raouafi2008}. Magnetic reconnection in polar regions has also been suggested as source of waves and energy to explain the fast wind \citep[e.g.][]{axford1999}. The polar rays observed in coronal maps of the wavelet power not necessarily are observed in coronal brightness maps, thus they might not always coincide with polar plumes and they are more extended than plumes. The same process of magnetic reconnection can in principle act at the base of plumes as well as rays, when they do not coincide. Hence, the lifetime of the density fluctuations derived in the present study (of the order of ten hours, middle right panel of Fig. \ref{fig:spatial_temporal_profiles_plume}) can provide information both on the duration of the intermittent phenomena and on the lifecycle of the plume itself, provided the same process is acting at the base of both kinds of structures. Moreover, the characteristic lifetime of the detected fluctuations could also be related to the timescale corresponding to the pronounced tendency of recurrence of plumes at the same locations \citep{lamy1997,deforest2001}.
\section{Conclusions}
This paper reports on the detection of transient periodic or quasi-periodic phenomena occurring in closed and open magnetic structures of the quiet solar corona out to $2\, R_{\odot}$. The technique used to reveal and characterize the fluctuations is based on the wavelet transforms, which represents a powerful tool to investigate the fine spatial pattern and temporal variability of fluctuating coronal features. Here, this technique has been used to map the variance of the total brightness of the outer solar corona from 1.4 to $4.0\,R_{\odot}$ for the first time and in a systematic way.

The results show that the solar corona is permeated by quasi-periodic recurrent brightness fluctuations, characterized by a wide range of spatial and temporal scales. They are however not uniformly distributed but clearly outline the coronal magnetic field, thus appearing as embedded in fluctuating magnetic structures. Moreover, they are significantly enhanced in regions of stronger field lines. In particular, it has been possible to identify an extended coronal arch in the mid-latitude equatorial region, likely part of a complex equatorial streamer present at the West limb, and a ray structure at high latitudes, both characterized by fluctuations of large amplitude. The arch-like structure is not observed in total brightness maps, thus implying that it is less dense than adjacent coronal structures. In the closed low and mid-latitude magnetic arch, the observed plasma fluctuations are consistent with the presence of slow magneto-acoustic waves trapped as standing modes and damped on about ten hours. These waves are very likely excited in a stochastic way by the convective motions at supergranular scales. The most likely mechanisms which could explain the wave damping are thermal conduction and phase mixing. In the polar region, the density plasma fluctuations of the ray structure, not necessarily coincident with a plume and even more extended than a plume, might be interpreted as signatures of intermittent, quasi-periodic magnetic reconnection due to continuously emerging flux tubes. This mechanism, considered the most probable one for the generation of plumes, as suggested in the literature, can in fact induce waves and ejections of mass; however, in order to explain the observed fluctuating feature, it should occur in a quasi-periodic recurrent way.

In conclusion, the present study of coronal density fluctuations, which extends previous analyses on periodic or quasi-periodic phenomena to cover the entire solar corona from the limb out to $4\,R_{\odot}$, provides new further information on the MHD coronal wave properties. The outer solar corona is a region where wave observations have been up to now highly fragmentary in space and time. The measurements of the power of MHD waves in the outer solar corona performed in this paper provide an accurate spatiotemporal reconstruction of quasi-periodic plasma phenomena. These may lead to a significant progress in determining how the energy carried by MHD waves is transferred from the photosphere to the corona, and there dissipated to accelerate the solar wind.
\begin{acknowledgements}
This work was supported by the Italian Space Agency (ASI) grants (I/023/09/0, I/013/12/0). The authors are thankful to the anonymous referee for helpful suggestions leading to a sounder version of the manuscript.
\end{acknowledgements}

\appendix
\section{Wavelet transforms and wavelet power spectrum in the analysis of the solar data}
\subsection{General formalism}
Given a wavelet function, $\Psi_{0}(\eta)$, depending on a non-dimensional 'time' parameter $\eta$, that has zero mean and is localized in both time and frequency space, the continuous wavelet transform of a $N$-point discrete two-dimensional time series $f(r,\phi,t)$, sampled at equal time spacing $\delta t$, where $r$ and $\phi$ are for instance the spatial polar coordinates and $t=0,...,(N-1)\delta t$ is the time, is the convolution of $f$ with a scaled and translated version of $\Psi_{0}(\eta)$ \citep{torrence1998}:

\begin{equation}
  W(r,\phi,t,s)=\frac{1}{\sqrt{N}}\sum_{t'=0}^{(N-1)\delta t}f(r,\phi,t')\Big(\frac{\delta t}{s}\Big)\Psi_{0}^{*}\Big(\frac{t'-t}{s}\Big),
  \label{eq:wavelet_transform}
\end{equation}

where the asterisk indicates the complex conjugate and $\Psi_{0}(\eta)$ is normalized to unit energy. By varying the wavelet scale parameter $s$ and translating the wavelet function $\Psi_{0}(\eta)$ in the time domain, it is possible to determine the amplitude of any periodic feature contained in the data and how this amplitude varies with scale and time. It is worth noting that the wavelet scale $s$ does not necessarily correspond to the Fourier period, $\tau$, because of the functional form of the wavelet function. The relationship between the equivalent Fourier period and the wavelet scale is derived analytically for a particular wavelet function by substituting a cosine wave of a known frequency $\nu=\tau^{-1}$ into Eq. \ref{eq:wavelet_transform} and computing the scale $s$ at which the wavelet power spectrum reaches its maximum.

There are a number of different wavelet functions $\Psi_{0}(\eta)$ with different properties that can be best suited to be applied to different studies. The best choice depends on first instance on its shape and width and whether it is a complex or real valued function. Complex wavelets are ideal candidates for capturing oscillatory behavior with respect to real wavelets, which are more useful to isolate peaks and discontinuities. The width of the wavelet function, namely the e-folding time of the wavelet power, is related to the resolution of the wavelet itself: a narrow function in time has good time resolution but poor frequency resolution and, viceversa, a broad function has poor time resolution and good frequency resolution. Two common mother wavelet functions, both complex, are the Paul and Morlet wavelets. The Paul wavelet is defined as

\begin{equation}
  \Psi_{0}(\eta)=\frac{2^{m}i^{m}m!}{\sqrt{\pi(2m)!}}(1-i\eta)^{-(m+1)},
  \label{eq:paul_wavelet}
\end{equation}

where $m$ ($=4$ in the present analysis) is the order of the wavelet. The width is $s/\sqrt{2}$ and the wavelet scale $s$ is related to the Fourier period $\tau$ by the expression $\tau=\frac{4\pi}{2m+1}s=1.4s$.

The Morlet wavelet consists instead of a plane wave modulated by a Gaussian:

\begin{equation}
  \Psi_{0}(\eta)=\pi^{-1/4}e^{i\omega_{0}\eta}e^{-\eta^{2}/2},
  \label{eq:morlet_wavelet}
\end{equation}

where $\omega_{0}$ ($=6$ in the present analysis) is the nondimensional frequency. The width is $\sqrt{2}s$ and the wavelet scale $s$ is related to the Fourier period $\tau$ by the expression $\tau=\frac{4\pi}{\omega_{0}+\sqrt{2+\omega_{0}^{2}}}s=1.03s$.

\begin{figure}
  \centering
  \includegraphics[height=\hsize,angle=90]{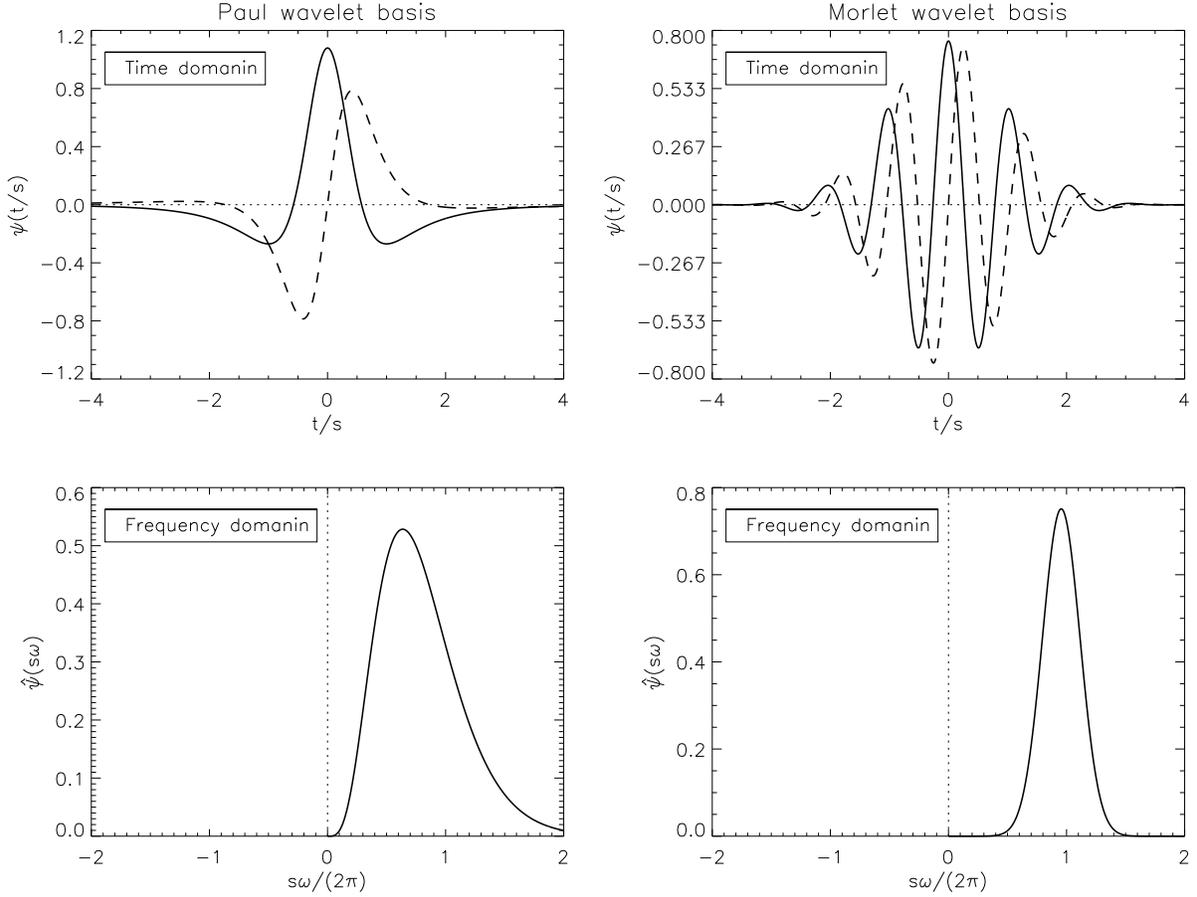}
  \caption{\emph{Top left}: Real part (solid) and imaginary part (dashed) of the Paul wavelet ($m=4$) in the time domain. \emph{Bottom left}: Paul wavelet in the frequency ($\omega$) domain. For plotting purposes, the scale $s$ was chosen to be $s=250\delta t$. \emph{Top right}: Real part (solid) and imaginary part (dashed) of the Morlet wavelet ($\omega_{0}=6$) in the time domain. \emph{Bottom right}: Morlet wavelet in the frequency ($\omega$) domain. For plotting purposes, the scale $s$ was chosen to be $s=250\delta t$ in both panels.}
  \label{fig:paul_morlet_wavelet}
\end{figure}

The Paul mother wavelet is more sharply defined in time in comparison to the Morlet function. Moreover it is characterized by few zero crossings and for this reason it can in principle identify many types of signatures in the power spectrum, as sharp variations in the input data and not just oscillatory modes. The Morlet function is narrower in periodicity space than the Paul wavelet, leading to a better frequency resolution. Moreover it contains more oscillations and then it is optimized for detecting oscillatory signals. Pictures of the real (solid lines) and imaginary (dashed lines) parts of the Paul and Morlet wavelets in both the time domain (top panels) and the frequency domain (bottom panels) are shown in Fig. \ref{fig:paul_morlet_wavelet}.

The wavelet power spectrum is defined as the square of the modulus of the wavelet transform $|W(r,\phi,t,s)|^{2}$. A vertical slice through the wavelet power spectrum is a measure of the spectrum at a given instant, so that the average of the wavelet power spectrum over the whole observational period, $0<t<(N-1)\delta t$, gives the global wavelet spectrum, defined as

\begin{equation}
  |\overline{W}(r,\phi,s)|^{2}=\frac{1}{N}\sum_{t=0}^{(N-1)\delta t}|W(r,\phi,t,s)|^{2}.
  \label{eq:global_wavelet_spectrum}
\end{equation}

It is worth noting that the global wavelet spectrum approaches the results that are obtained with the Fourier spectrum of the considered time series \citep{percival1995}.

The average wavelet power, that is the sum of the power over all the considered periodicities between $s_{1}$ and $s_{2}$, is given by:

\begin{equation}
  |\overline{W}(r,\phi,t)|^{2}=\frac{\delta j\delta t}{C_{\delta}}\sum_{j=j_{1}}^{j_{2}}\frac{|W(r,\phi,t,s_{j})|^{2}}{s_{j}},
  \label{eq:scale_averaged_wavelet_power}
\end{equation}

where $\delta j$ (=0.125 in the present analysis) is the spacing between the discrete scales commonly defined as fractional powers of two, $s_{j}=s_{0}2^{j\delta j}$ ($s_{0}$ is the Nyquist-Shannon scale), and $C_{\delta}$ is the reconstruction factor for the assumed wavelet function (we adopted 1.132 and 0.776 for the Paul and Morlet wavelets, respectively).

The scale-averaged wavelet power defined by Eq. \ref{eq:scale_averaged_wavelet_power} gives a measure of the average variance of the time series as a function of time, over the range of periodicities considered in the analysis.
\subsection{Cone of influence}
Because of the finite length of the time series, errors will arise at the beginning and end of the wavelet power spectrum, as the Fourier transform assumes the data are cyclic. The Cone Of Influence (COI) is defined as the region of the wavelet spectrum where edge effects become important and is commonly defined as the e-folding time for the autocorrelation of the wavelet power at each scale. This e-folding time is chosen so that the wavelet power for discontinuity at the edge drops by a factor $e^{-2}$ and assures that the edge effects are negligible beyond this point.
\subsection{Significance levels}
Determining significance levels for a wavelet spectrum needs to choose an appropriate background spectrum in order to establish a null hypothesis against which the inferred spectrum must be tested. If a peak in the wavelet power spectrum is significantly above the background spectrum, it can be assumed to be a true feature with a certain percent confidence. In many astrophysical applications it is a good choice to adopt as a background spectrum a flat spectrum resulting from wavelet analysis of pure white-noise, normally distributed time series. The wavelet coefficients of a normally distributed random variable are normally distributed too; the square of a normally distributed variable is chi-square distributed with one degree of freedom, $\chi_{1}^{2}$. Hence, in the hypothesis of white-noise normally distributed time series, the wavelet power spectrum would be $\chi_{2}^{2}$ distributed with two degrees of freedom. The 99\% confidence contour lines are then determined by multiplying the background spectrum by the 99th percentile value of $\chi_{2}^{2}$, that is $\chi_{2}^{2}(99\%)$.
\end{document}